# Requirements Framing Affects Design Creativity

Rahul Mohanani, Burak Turhan, *Member, IEEE* and Paul Ralph

**Abstract**—Design creativity, the originality and practicality of a solution concept is critical for the success of many software projects. However, little research has investigated the relationship between the way desiderata are presented and design creativity. This study therefore investigates the impact of presenting desiderata as ideas, requirements or prioritized requirements on design creativity. Two between-subjects randomized controlled experiments were conducted with 42 and 34 participants. Participants were asked to create design concepts from a list of desiderata. Participants who received desiderata framed as requirements or prioritized requirements created designs that are, on average, less original but more practical than the designs created by participants who received desiderata framed as ideas. This suggests that more formal, structured presentations of desiderata are less appropriate where a creative solution is desired. The results also show that design performance is highly susceptible to minor changes in the vernacular used to communicate desiderata.

✦

## 1 INTRODUCTION

THE software engineering (SE) research community widely accepts that understanding system requirements is critical for designing good software systems [1]. The assumptions that software projects have discoverable and documentable requirements, and that good requirements specifications lead to good software [2], have stimulated diverse attempts to improve RE processes. For example, prioritizing requirements is gaining popularity in complex software-intensive projects [3]–[5].

One might expect a clear, well-structured, simplified and rationalized account of what is *required* to increase the odds of a good system design [6], especially in mission/safety critical domains. However, task structure is negatively associated with design performance [7]. Specifically, "over-concentration (over-structuring) on problem definition does not necessarily lead to successful design outcomes" [8, p. 439]. In other words, labelling desiderata as requirements may reduce design performance, especially in non-critical application domains where creativity and innovation makes the business difference. Yet, SE research has not empirically investigated the effects of presenting (framing) desiderata in different ways. Faced with explicit requirements, practitioners tend to fixate on early solution ideas or existing solutions and produce less original designs [9].

This raises an interesting question. Suppose we have a list of things that are needed or wanted for some system. Does the way we present that list to a product designer affect their performance? More formally:

**Research Question:** *Does the framing of desiderata affect design creativity?*

Here, a *desideratum* is a property of a real or imagined system that is wanted, needed or preferred by one or more project stakeholders. Framing alludes to the *framing effect*:

"the tendency to give different responses to problems that have surface dissimilarities but are formally identical" [10, p. 88]. Here, it specifically refers to the way in which desiderata are presented (e.g. a list of "the system shall..." specifications [11], a backlog of user stories [12], a collection of use case narratives [13]). Meanwhile, creativity is a cognitive process of generating one or more ideas (in this case, for a software artifact) that are not only novel but also feasible [14]. Consequently, we operationalize *design creativity* in terms of both originality and practicality.

This article extends an earlier study investigating the effect of desiderata framing on the originality of design concepts [15]. This earlier study found that labelling desiderata as requirements led to less original designs. Several colleagues suggested that originality is only one aspect of design performance and that prioritizing the requirements should improve performance. We were therefore inspired to perform a second study with prioritized requirements, and to evaluate practicality as well as originality of designs.

This article therefore extends our earlier work in two ways. First, we re-analyze the data from our previous study to measure the effect of the treatment on the practicality of the solution concepts. Second, we conduct another experiment to investigate the effects of presenting desiderata as prioritized requirements on both originality and practicality. This paper also includes a significantly updated discussion. Parts of this paper, including sections 2 and 4, re-use text from the earlier publication, where relevant.

Next, we review existing literature on fixation, the framing effect and creativity (Section 2). Then, we describe the research method (Section 3), followed by the analysis and results (Section 4). Section 5 interprets the results and summarizes the study's implications. Section 6 discusses the threats to validity. Section 7 concludes the paper with a summary of its contributions.

## 2 RELATED WORK

This section provides an overview of the major concepts that contribute to the theory behind the two experiments.

- R. Mohanani is with the Dept. of CSE & HCD, IIIT Delhi, India.
  E-mail: rahul.mohanani@iiitd.ac.in
- B. Turhan is with the Faculty of IT, Monash Univ., Australia.
  E-mail: burak.turhan@monash.edu
- P. Ralph is with the Dept. of Comp. Sci., Univ. of Auckland, New Zealand.
  E-mail: paul@paulralph.name

## 2.1 Fixation and the Framing Effect

Cognitive biases are systematic deviations from optimal reasoning [16]. They help to explain many common problems in diverse SE activities including design (e.g. [17]), testing (e.g. [18]), requirements engineering (e.g. [19]), project management (e.g. [20]) and in SE generally (e.g. [21]). Research on cognitive biases is useful not only to identify common errors, their causes and how to avoid them in SE processes, but also for developing better practices [22], methods [23], and artifacts [24]. The current study is primarily concerned with two cognitive biases—*fixation* and *the framing effect*.

The way decision options are presented biases decision making [25]. Specifically, the *framing effect* is "the tendency to give different responses to problems that have surface dissimilarities but that are really formally identical" [10, p. 88]. For example, in one experiment, participants were asked to choose between two treatments for a hypothetical disease—treatment A would save 200 of 600 people, whereas, treatment B had 1 in 3 chance of saving everyone and 2 in 3 chance of saving no one. Participants were asked to choose between 400 people definitely dying or a 1 in 3 chance that no one will die. Although the expected effectiveness of the two treatments are exactly the same, most of the participants chose the latter. However, when participants were asked to choose between definitely saving 200 people or a 1 in 3 chance of saving everyone, most chose the former [26]. Here, the difference in response is determined by the way the question is framed rather than the facts. The framing effect is extremely robust [27], affects many individuals across diverse circumstances, and has not been studied much in SE.

*Fixation*, originally proposed by Freud regarding human sexuality, gradually broadened to refer to a tendency to "disproportionately focus on one aspect of an event, object or situation, especially self-imposed or imaginary obstacles" [28, p. 5]. Fixation is one of many biases related to selective attention—the tendency for different people to perceive the same events differently [29]. Several experiments have demonstrated *design fixation*—the tendency for designers to generate solutions very similar to given examples [30], [31] or existing artifacts [32]. In other words, fixation prevents software designers from making original or novel associations, or envisioning alternative solutions [33].

Providing design examples causes design fixation, but the effects of design fixation are moderated by several factors; for example: (1) the propensity for fixation varies by domain; for instance, mechanical engineers fixate more than industrial designers [34]; (2) common examples induce greater fixation than unusual examples [35]; (3) priming designers with good examples leads to better design performance than priming designers with intentionally flawed examples or no examples [36]; (4) Fixation is partially determined by the way the task is framed [33]. Fig. 1 summarizes these effects.

This last point highlights the relationship between fixation and framing. Earlier work (e.g. [37], [38]) has investigated fixation by giving participants different instructions, where the independent variable was task framing and the resulting fixation were conceptualized as a kind of framing effect. Of course, individuals can become fixated without any experimental framing intervention. However, many studies on fixation leverage framing effects (e.g. [39], [40]). Consequently, instead of *designers fixate on given examples* we can think of these studies as showing that *task framing causes fixation.* (Below, we also use framing to induce fixation.)

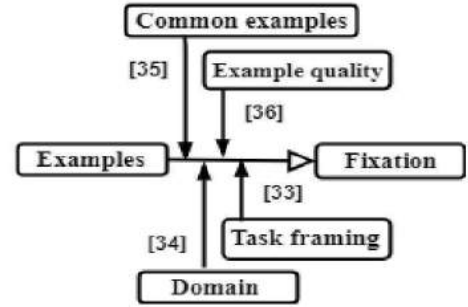

Fig. 1. Model of Design Fixation
*Note: Filled arrows indicate moderating effect; unfilled arrows indicate causation.*

## 2.2 Creativity

No single definition of creativity is universally accepted; however, for our purposes creativity refers to "production of novel and useful ideas by an individual or small group of individuals working together" [41, p. 127]. Creativity is often linked with divergent thinking [42]—exploring many diverse solutions to a problem.

Creativity is a multi-dimensional construct [41]; a creative artifact is not only *novel* [43], [44] but also *practically useful* [14], [45]–[47]. This dual criteria view of creativity is widely accepted in engineering design (e.g. [48]) and specifically software engineering (e.g. [49]). However, these criteria are not equal—when assessing creativity, we consider usefulness only if the solution is novel [50].

The dual criteria view suggests assessing creativity by evaluating an artifact's originality and practicality [51], [52]. However, since no objective metrics for originality or practicality exist, studies typically use expert judges to assess creativity [53], [54], [55]. (Below, we also use expert judges to assess creativity.)

SE research predominantly considers creativity in the context of requirements engineering [56], software design [57], agile development [58] and open source software [59].

The relationship between creativity and RE is highly controversial. Many scholarly articles, textbooks and official standards present requirements engineering as *discovering* requirements in an objective reality and inferring system properties from these objective requirements (cf. [60]). In contrast, many scholars have proposed applying creativity techniques to requirements engineering, including creativity workshops, brainstorming, mind-mapping, the 5W1H method (i.e. asking when, where, who, what, why and how to gather new ideas), and various interactive collaboration techniques [57], [61], [62]. Some scholars consider RE as an intrinsically creative process [61], [63] where software requirements encapsulate the results of creative thinking about the system [49].

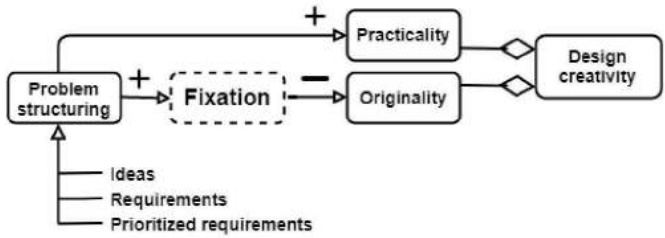

Fig. 2. Theoretical model for requirements fixation

*Note: Unfilled arrows indicate causation; unfilled diamond indicates aggregation; plus and minus signs indicate magnitude of effect.*

## 3 RESEARCH METHOD

The section describes two between-subjects, randomized controlled experiments, Experiment One and Experiment Two, that investigate the relationship between desiderata framing and design creativity (see Table 1 for an overview). In both experiments, participants with some software design experience were randomly assigned to a treatment group or control group. All participants were given the same set of desiderata and asked to design, on paper, a mobile application. The designs were then shuffled and assessed by two experts.

Experiment One compares desiderata framed as requirements (treatment) to desiderata framed as ideas (control); Experiment Two compares prioritized requirements (treatment) to ideas (control). We chose ideas framing against requirements because an *idea* could be either good or bad, promotes skepticism and generally does not connote any sort of preference. Moreover, we conceptualized ideas framing as a simple yet tractable alternative to minimize the difference between the two groups and to isolate the psychological effects of framing desiderata as requirements. We continued to use ideas framing in the second experiment to maintain maximum control over cross-study synthesis and to maximize the validity of the results by combining the two experiments (see Section 4.3).

### 3.1 Hypotheses

More abstract, uncertain, incomplete problem framing leads to solutions that are more original than more specific, structured problem framing does, for at least three reasons:

1) More structure encourages fixation on early solution ideas [64], established solutions [30], [34] and bogus requirements [15].
2) Non-specific or open goals lead to more learning [65] by reducing cognitive load [66] and increasing unconscious assimilation of problem relevant information [67].
3) Presenting conflicting objectives leads to more creative solutions [68].

In other words, presenting a task with more ambiguity and uncertainty leads to more exploration, learning, and departing from norms (i.e. divergent thinking). This leads to designs that are more original but sometimes less practical.

Presenting desiderata as *ideas* is more ambiguous and uncertain than presenting desiderata as *requirements*. Here,

the designer may wonder, *what are these ideas?* and *how should they be used?* Prioritizing the requirements adds further structure and clarity. We therefore hypothesize that participants who received more structured desiderata framing will produce designs that are less original and more practical (see Fig. 2); more formally:

$H_1$: *Participants who receive desiderata framed as requirements will produce design concepts that are less original than the design concepts produced by participants who received desiderata framed as ideas.*

$H_2$: *Participants who receive desiderata framed as requirements will produce design concepts that are more practical than the design concepts produced by participants who received desiderata framed as ideas.*

$H_3$: *Participants who receive desiderata framed as prioritized requirements will produce design concepts that are less original than the design concepts produced by participants who received desiderata framed as ideas.*

$H_4$: *Participants who receive desiderata framed as prioritized requirements will produce design concepts that are more practical than the design concepts produced by participants who received desiderata framed as ideas.*

In other words, Experiment One tests whether desiderata framed as requirements leads to designs that are more practical but less original than desiderata framed as ideas. Experiment Two tests whether desiderata framed as prioritized requirements leads to designs that are more practical but less original than desiderata framed as ideas.

### 3.2 Materials and procedure

We compiled a list of 25 desiderata for a health and fitness mobile application based on existing applications in this genre. The idea was to create a manageable list of desiderata that participants might be familiar (but not well-versed) with. As our intended participants were not trained RE professionals or expert designers (see Section 3.4), we aimed for a realistic, yet imperfect specification document that might be created by a client without formal RE training, rather than the polished work of an expert requirements analyst [69]. Since we could not find a similar available list, we compiled the desiderata based on features of existing health-fitness apps for Experiment One. We decided to use the exact same list, albeit prioritized, for Experiment Two to make cross-study comparisons more valid (as explained in Section 3) and to maintain a consistent series of experiments [70].

We created three versions of the list. Each version contained the same desiderata (e.g. "measure calorie intake") in the same order. The only differences were: in the first version, the desiderata were called *ideas* and phrased "The system might measure calorie intake"; in the second version, the desiderata were called *requirements* and phrased "The system shall measure calorie intake"; and in the third version, the desiderata were organized into five priority levels. The priority levels were determined by asking seven experienced colleagues from the first author's university to prioritize the desiderata by importance. We used a weighted average to combine their judgments.

TABLE 1
Overview of experimental design

| Design element | Experiment One | Experiment Two |
|---|---|---|
| Location | United Kingdom | Finland |
| Number of participants | 42 | 34 |
| Mean age of participants (years) | 25 | 26 |
| Treatment group | Desiderata framed as requirements | Desiderata framed as prioritized requirements |
| Control group | Desiderata framed as ideas | |
| Dependent variables | Design originality and practicality | |
| Dependent variable scale | Ordinal | |
| Instrumentation | Expert judgment | |
| Main statistical test | Mann Whitney U test (non-parametric) | |

Each experiment was conducted in two parallel sessions, in separate rooms with very similar dimensions and settings. Participants were randomly assigned to one of the two rooms on arrival by the invigilators. Participants signed a consent form and then filled in a pre-task (demographic) questionnaire. Next, the invigilators distributed the task documents: the list of desiderata and identical design templates comprising several blank mobile screen-sized boxes in portrait and landscape orientations, with space for written explanations. In both experiments, the control group received the ideas document. In Experiment One, the treatment group received the requirements document; in Experiment Two, the treatment group received the prioritized requirements document. Participants were given 60 minutes to complete the design task, after which they filled our a post-task questionnaire including a manipulation check (described below). All of the task documents are available in our replication package [1].

### 3.3 Data collection and assessment procedure

The designs were de-identified, combined into a single set, randomly ordered and given to two independent, expert judges. The first author and a colleague who is a senior researcher in software design and creativity evaluated the design concepts for Experiment One. For Experiment Two, a highly recognized expert (professor) in digital design for public health-care sector in the first author's university and a software professional with 8 years of experience in software design and development for large scale ERP systems were the two judges. As in similar studies (e.g. [71]–[73]), the judges evaluated the designs using the Consensual Assessment Technique [74]. That is, the judges scored each design for originality and practicality on a five-point scale from low (1) to high (5). Here *originality* means being new and not found or implemented in similar applications, while *practicality* means being straightforward and appropriate for an actual use in a specific context, rather than a hypothetical use.

For each experiment, the judges discussed the meaning of originality and practicality, then graded three randomly selected designs together to establish a common understanding of the grading procedure. The judges then evaluated the remaining designs independently. Disagreements were resolved by a third expert judge. The data analysis script and final grades are also available in our replication package [1]. Fig. 3 and Fig. 4 are examples of a highly original and a highly practical design respectively. Both, Fig. 3 and Fig. 4, represents the 'log-in screen' for the mobile application.

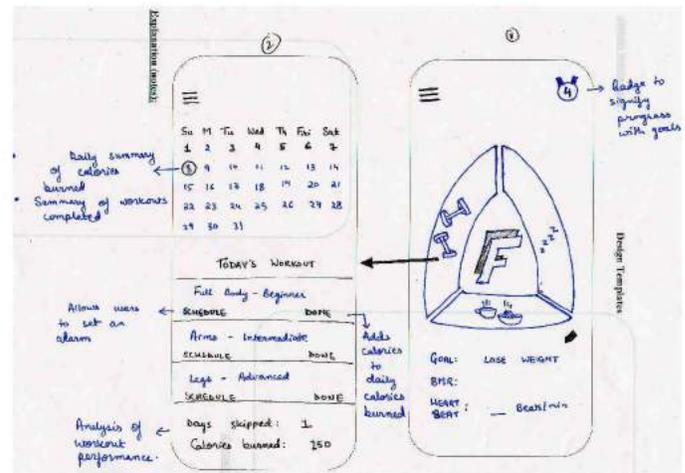

Fig. 3. Example of a highly original design

### 3.4 Participants

For Experiment One, participation was solicited from management and engineering postgraduate students enrolled at Lancaster University, United Kingdom using the relevant student mailing lists. A convenience sample of 19 females and 23 males with a mean age of 25 years (standard deviation 6.07) participated. Fourteen participants had a software engineering background. Participants received a complimentary lunch coupon for participating in the study.

For Experiment Two, participants were recruited from postgraduate students enrolled in the information processing science program at the University of Oulu, Finland. A convenience sample of 34 participants were selected, comprising of seven female and 27 male subjects with an average age of 26 years (standard deviation 5.83). Participants received extra credit as a part of their coursework.

All of the participants had at least one year of experience in software development. No participants had experience

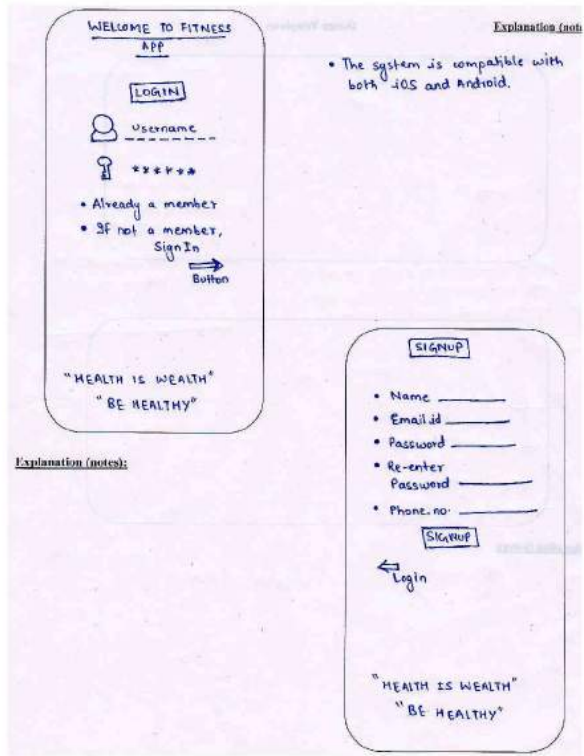

Fig. 4. Example of a highly practical design

in the specific task domain of health and fitness mobile applications.

## 4 ANALYSIS AND RESULTS

This section summarizes the results of the two experiments, followed by a cross-study synthesis, which combines the data from both the experiments.

### 4.1 Experiment One

#### 4.1.1 Reliability

Participants produced 42 designs. The judges agreed on the originality of 34 and the practicality of 36 designs. This gives inter-rater agreement (Cohen's kappa) of 0.67 for originality and 0.78 for practicality grades, which indicates 'substantial agreement', and hence acceptable reliability [75].

#### 4.1.2 Hypothesis testing

To test Hypotheses $H_1$ and $H_2$, we compared the distributions of originality (Table 2; Fig. 5) and practicality (Table 3; Fig. 6) grades across the two groups. Both, originality (p=0.08) and practicality (p=0.153) grades meet the assumption for homogeneity of variance (Levene's nonparametric test [76]). However, neither originality nor practicality grades are normally distributed (Shapiro-Wilk test [77] p=0.003 for both). The data therefore meets the four assumptions of the Mann-Whitney U test:

- The dependent variable is ordinal or continuous.
- The independent variable has two categorically independent groups.
- Independence of observations (e.g. a between-subjects design).
- The dependent variable exhibits homogeneity of variance.

*Hypothesis $H_1$:* Participants who received the ideas framing produced designs that are significantly more original (U=116.5; n=42; p=0.004). The effect size (Cliff's $\delta$= -0.49)[2] indicates a 'high-effect' with 95% CI [-0.74, -0.13] [79].

*Hypothesis $H_2$:* In contrast, participants who received desiderata framed as *requirements* produced designs that are significantly more practical (U=128.5; n=42; p=0.018). The effect size (Cliff's $\delta$=0.41) indicates a 'medium-high' effect with 95% CI [0.06, 0.67] [79].

#### 4.1.3 Exploratory analysis

In Section 3.1, we theorized that framing desiderata as requirements would increase designers' propensity for fixation. While thoroughly investigating the cognitive mechanisms underlying fixation would require a more exploratory kind of study (e.g. a think-aloud protocol study [80]), the post-task questionnaire included a simple manipulation check. The participants were asked to rate, on a five-point scale, the importance of the list of desiderata (as requirements or ideas) for guiding their designs. If the manipulation was successful:

- the treatment group (requirements framing) should rate the desiderata as more important than the control group (ideas framing);
- desiderata importance should be directly related to practicality;
- desiderata importance should be inversely related to originality.

The treatment group (median=4) rated the desiderata as more important than the control group (median=3). The statistical significance of this difference is difficult to assess because the data exhibits neither normality, as assumed by the independent samples t-test; nor homogeneity of variance, as assumed by the Mann-Whitney U test. However, both tests suggest that the difference is statistically significant (n=42; U=125.5 p=0.011; t=2.97, p=0.006). The effect size (Cliff's $\delta$=0.43) indicates a 'medium-high' effect with a 95% CI [0.09, 0.68].

Meanwhile, desiderata importance is directly related to practicality (Spearman correlation, rho=0.193) and inversely related to originality (rho=-0.205) but neither difference is statistically significant (p=0.111, p=0.096 respectively) [3].

### 4.2 Experiment Two

#### 4.2.1 Reliability

The participants produced 32 total designs. The judges agreed on 26 originality scores (Cohen's kappa=0.7) and 23 practicality scores (Cohen's kappa=0.65). This indicates 'substantial agreement' [75] and therefore acceptable reliability.

---

2. Cliff's Delta ($\delta$) for the Mann-Whitney U test is calculated by using the equation: $\#(X_1 > X_2) - \#(X_1 < X2)/n_1 n_2$, where $X_1$ and $X_2$ are scores within the groups (i.e. Group A and Group B) and $n_1$ and $n_2$ are the sizes of the sample groups. The cardinality symbol # indicates counting [78].

3. All of the correlations reported in this paper are one-tailed because we have an a priori theoretical reason to expect: (1) direct correlations between problem structure and practicality; and (2) inverse correlations between problem structure and originality.

TABLE 2
Frequency of originality grades (Experiment One)

| Grade | Treatment (Requirements) | Control (Ideas) |
|-------|--------------------------|-----------------|
| 1 | 1 | 2 |
| 2 | 7 | 1 |
| 3 | 12 | 6 |
| 4 | 0 | 10 |
| 5 | 1 | 2 |
| Mean | 2.67 | 3.43 |
| Median | 3 | 4 |

TABLE 3
Frequency of practicality grades (Experiment One)

| Grade | Treatment (Requirements) | Control (Ideas) |
|-------|--------------------------|-----------------|
| 1 | 2 | 5 |
| 2 | 2 | 6 |
| 3 | 6 | 5 |
| 4 | 6 | 4 |
| 5 | 5 | 1 |
| Mean | 3.48 | 2.52 |
| Median | 4 | 2 |

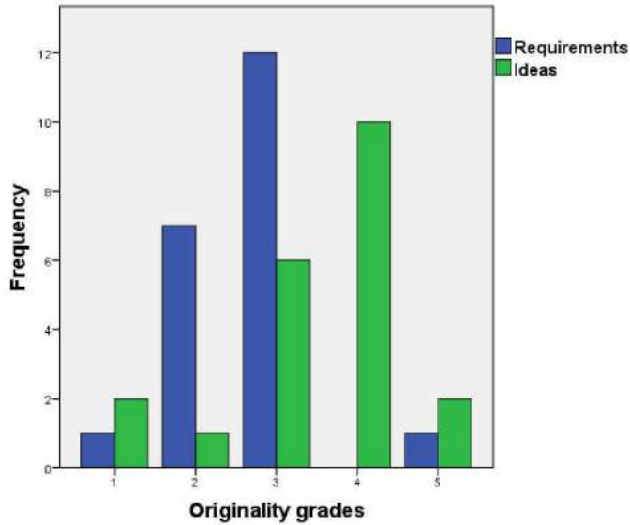

Fig. 5. Originality grades across both the groups (Experiment One)

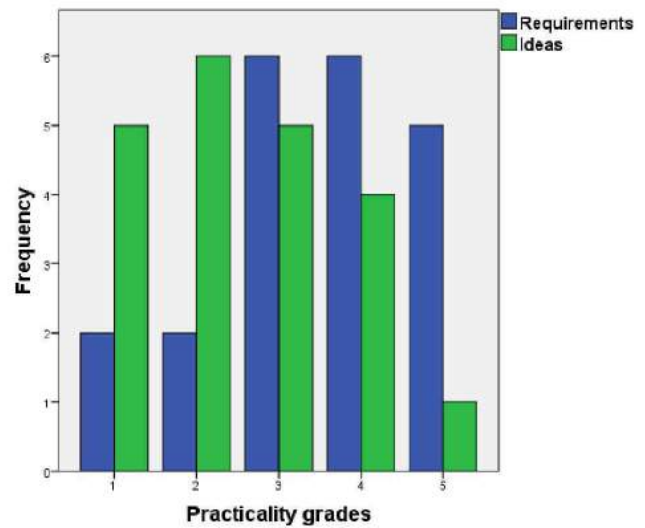

Fig. 6. Practicality grades across both the groups (Experiment One)

### 4.2.2 Hypothesis testing

To test hypotheses $H_3$ and $H_4$, we compared the distributions of originality (Table 5; Fig. 7) and practicality (Table 6; Fig. 8) scores across the two groups. As in Experiment One, the dependent variables exhibit homogeneity of variance—originality (Levene's test p=0.319); practicality (p=0.138)—but not normality—originality (Shapiro-Wilk test p=0.02); practicality (p<0.001). We therefore again employ the Mann-Whitney U test.

*Hypothesis $H_3$:* Participants who received the ideas framing produced designs that are significantly more original (Mann-Whitney U test; U=84, n=34, p=0.02). The effect size (Cliff's $\delta$= -0.43); 95% CI [-0.71, -0.05] indicates a 'medium-high' effect.

*Hypothesis $H_4$:* In contrast, participants who received desiderata framed as *prioritized requirements* produced designs that are significantly more practical (Mann-Whitney U test; U=77, n=34, p=0.02). The effect size (Cliff's $\delta$=0.46); 95% CI [0.07, 0.73] indicates a 'medium-high' effect.

### 4.2.3 Exploratory analysis

As in Experiment One, we included the desiderata-importance manipulation check in the post-study questionnaire. Again, participants in the treatment group (prioritized requirements) indicated that the desiderata were more important than participants in the control group, with medians of 4 and 2 respectively (Table 7). Again, the data is non-normal with unequal variance, but both the Mann-

Whitney U test and independent samples t-test suggest that the difference is statistically significant for importance of desiderata (n=34; U=36.5, p<0.001; t=4.92, p<0.001) with a 'high' effect size (Cliff's $\delta$=0.74; 95% CI [0.38, 0.91]).

Again, desiderata importance is directly related to practicality (Spearman correlation, rho=0.374; p=0.015) and inversely related to originality (rho=-0.195; p=0.135) but this time the relationship between perceived desiderata importance and the practicality of the resulting design concepts is statistically significant.

Participants were also asked, *"How confident are you that the conceptual design(s) you created can be implemented as a mobile app?"* with a five-point scale from 'not sure' to 'extremely sure' (Table 8). The treatment group—in our case, *prioritized requirements* group, indicated greater confidence in the practicality of their designs than the control group

TABLE 4
Frequency of 'importance of desiderata' scores (Experiment One)

| Importance | Treatment (Requirements) | Control (Ideas) |
|------------|--------------------------|-----------------|
| 1 (least) | 0 | 3 |
| 2 | 1 | 5 |
| 3 | 2 | 3 |
| 4 | 12 | 6 |
| 5 (most) | 6 | 4 |
| Mean | 4.10 | 3.14 |
| Median | 4 | 3 |

#### TABLE 5
Frequency of originality grades (Experiment Two)

| Grade | Treatment (Prioritized req.) | Control (Ideas) |
|---|---|---|
| 1 | 6 | 2 |
| 2 | 6 | 5 |
| 3 | 3 | 2 |
| 4 | 2 | 6 |
| 5 | 0 | 2 |
| Mean | 2.06 | 3.06 |
| Median | 2 | 3 |

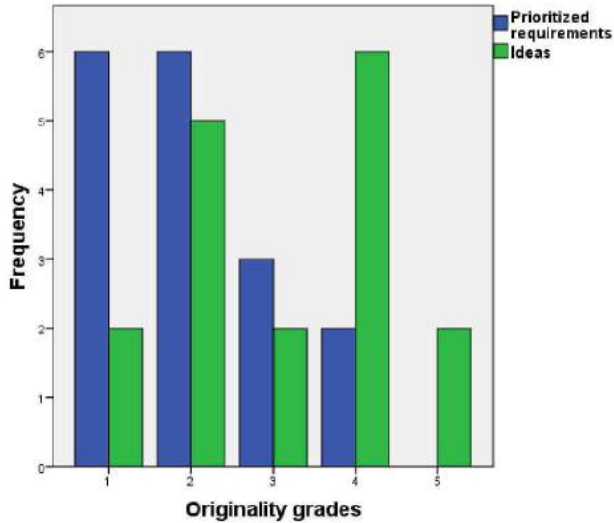

Fig. 7. Originality grades across both the groups (Experiment Two)

#### TABLE 6
Frequency of practicality grades (Experiment Two)

| Grade | Treatment (Prioritized req.) | Control (Ideas) |
|---|---|---|
| 1 | 4 | 8 |
| 2 | 2 | 3 |
| 3 | 2 | 5 |
| 4 | 6 | 1 |
| 5 | 3 | 0 |
| Mean | 3.12 | 1.94 |
| Median | 4 | 2 |

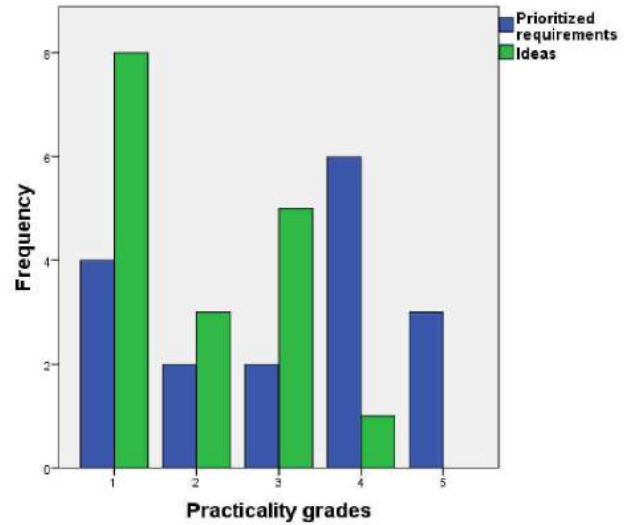

Fig. 8. Practicality grades across both the groups (Experiment Two)

(U=77.5, n=34, p=0.017; t=2.60, p=0.014) with a 'medium-high' effect size (Cliff's $\delta$=0.46; 95% CI [0.07, 0.73]). Participants responses were positively correlated with the judges' practicality scores, but the correlation was not quite significant (rho=0.284; p=0.052).

### 4.3 Cross-study synthesis

To conduct a meta-analysis, we combined the data from both experiments into three groups—the control group (ideas; n=38), treatment one (requirements; n=21), and treatment two (prioritized requirements; n=17)—producing a clear pattern (Table 9). More structure leads to less originality (rho=0.430, p<0.001). Both treatment groups exhibit better practicality than the control group (Kruskal-Wallis H=11.978, p=0.003) [4].

Prioritizing the requirements counter-intuitively appears to reduce practicality, but this result should be interpreted with caution because it involves comparing unequal groups from two different studies.

Does fixation mediate the relationship between problem structure and creativity? Mediation requires four conditions:

1) The independent variable is correlated with the dependent variable. (Supported: rho=-0.430, p<0.001 for originality; rho=0.338, p=0.001 for practicality.)
2) The independent variable is correlated with the mediating variable. (Supported: rho=0.461, p<0.001.)

#### TABLE 7
Frequency of 'importance of desiderata' scores (Experiment Two)

| Importance | Treatment (Prioritized req.) | Control (Ideas) |
|---|---|---|
| 1 (least) | 0 | 5 |
| 2 | 1 | 5 |
| 3 | 2 | 5 |
| 4 | 11 | 1 |
| 5 (most) | 3 | 1 |
| Mean | 3.94 | 2.29 |
| Median | 4 | 2 |

#### TABLE 8
Frequency of 'confidence of implementation' scores (Experiment Two)

| Confidence | Treatment (Pri. req.) | Control (Ideas) |
|---|---|---|
| 1 (not sure) | 1 | 2 |
| 2 | 0 | 6 |
| 3 | 3 | 2 |
| 4 | 7 | 5 |
| 5 (extremely sure) | 6 | 2 |
| Mean | 4.00 | 2.94 |
| Median | 4 | 3 |

4. Again, the data exhibits equal variances (Levene's non-parametric test, p=0.06 for originality; p=0.25 for practicality) grades but is not normally distributed (Shapiro-Wilk test for normality confirmed non-normal distributions for both originality (p<0.001) and practicality (p<0.001) grades.

TABLE 9
Cross-study synthesis

|  | Control (Ideas) | Treatment 1 (Requirements) | Treatment 2 (Prior. req.) |
|---|---|---|---|
| mean originality | 3.26 | 2.67 | 2.06 |
| mean practicality | 2.26 | 3.48 | 3.12 |
| n | 38 | 21 | 17 |

3) The mediating variable is correlated with the dependent variables. (Partially supported; importance of desiderata is positively correlated with practicality—rho=0.297, p=0.005—and negatively correlated with originality—rho=-0.144, p=0.108); however, the latter is not statistically significant).

4) Controlling for the mediating variable reduces the absolute value of the correlation between the independent and dependent variables. (For practicality rho drops from 0.338 to 0.201; for originality, from -0.430 to -0.400.)

This suggests that fixation mediates the effect of problem structure on practicality. It could mean that fixation does not mediate the effect of problem structure on originality, or it could mean that this effect is not very strong and our sample was not large enough to detect it. In either case, this is not what we expected. Our understanding of the literature suggests that fixation mediates the relationship between problem structure and originality, not practicality. We have no theoretical justification for benefiting practicality.

## 5 DISCUSSION

The needs, desires, preferences, and conjectures of project stakeholders can be framed in many ways—as requirements, user stories, scenarios, use cases, or simply as ideas. Prior research suggests that more structured, unambiguous task framing leads to designs that are less original but more practical (see Section 2). However, no previous studies have established whether this phenomenon applies to software engineering.

The purpose of the two experiments reported above is therefore to investigate whether a small change in the way desiderata are framed can induce a significant difference in design outcomes. Each experiment indicates that presenting the same collection of desiderata in slightly different ways has a profound effect on design performance. Presenting desiderata as requirements leads to significantly designs that are more practical but less original.

The meaning of this finding is clear. When a project calls for innovation—significant departures from current approaches—desiderata should not be presented as requirements. In contrast, when a project calls for more straightforward, archetypal solutions, presenting desiderata as requirements is more appropriate. Being randomized controlled trials, these findings are very robust, and have well-understood limitations (see Section 6).

These findings conflict with the dominant view in requirements engineering (RE), namely, that providing more clarity, precision and structure is critical for success. Some RE experts may struggle to accept that more ambiguity is sometimes better. We think this is rooted in the different worldviews of requirements analysts (or at least, the requirements engineering academic literature) and designers (or at least, the interdisciplinary literature on design).

In RE, a requirement is something that is *demanded*, because success means delivering whatever was agreed. If the client wants the funeral home's website all in Comic Sans font, then that is a requirement. In user-centered design, a requirement is something that is *needed*, because success means that the stakeholders benefit from the system. The funeral home's website will be in an appropriately sober, presumably not Comic Sans font, because it is better for both the users and the business. Moreover, one may further argue for a distinction between *mandatory* and *optional* requirements. However, the psychological effects of mislabeling ostensible features of a system as requirements suggests that using any additional labels will likely increase confusion and curtail creativity [81].

In RE, it is often assumed that users have requirements—the trick is eliciting them. Empirical research does not support this view. People generally do not answer questions by retrieving relevant enduring preferences from an internal directory. Rather, during an interview, the analyst and the user co-construct evanescent preferences [82]. Although many RE researchers have suggested that requirements are invented from a set of conflicting opinions, beliefs, wishes or preferences (e.g. [63], [83]), yet much RE research still continues to assume that requirements are elicited. Labeling a fleeting preference as a *requirement* inflates both its estimated importance and our confidence in that estimate. Given such requirements to a designer, paints an inaccurate picture of the project context.

This brings us to the results of our exploratory analysis. While this analysis has more threats to validity (see Section 6), it does suggest a strong negative relationship between problem-structuring and design originality. The obvious question is, why? *Why would designers produce design concepts that are less original and more practical in response to more formal desiderata framing?* Although the main goal of our experiments was not to answer this question, we can make some educated guesses.

Previous research on design fixation investigated how providing designers with explicit examples reduces their creativity [32], [33]. We extend this idea by showing that framing desiderata as requirements similarly diminishes originality, even without examples. This is also similar to mental-set fixation—a situation where practitioners restrict the use of their own abilities due to situationally induced bias [30]. Meanwhile, some designers are biased against originality a priori (cf. [84], [85]) and expert designers tend to treat given desiderata and constraints more skeptically regardless of framing [86].

In other words, we suspect that the high importance and confidence connoted by the term *requirement*, coupled with a predetermined order of implementation (i.e. prioritized requirements), shuts down participants' creative potential by promoting the view that the problem is already well-understood and largely solved. The requirements framing induces participants to fixate on the stated desiderata instead of exploring the solution space.

This led us to theorize that software professionals are prone to *requirements fixation*—the tendency to rely heavily on desiderata that are explicitly framed as requirements. Some of the symptoms of requirements fixation may include:

- Failing to question doubtful, ambiguous or conflicting desiderata.
- Perceiving desiderata as having equal (high) importance.
- Perceiving desiderata as having equal confidence.
- Failing to consider the difference between project goals and desiderata.
- Failing to consider implicit or non-functional desiderata.

Our exploratory analysis only partially supports requirements fixation. Fixation mediated the relationship between framing and practicality, but not originality—all the correlations are in the expected directions but the correlation between fixation and originality is small (by Cohen's standard) and not statistically significant. More research on this is clearly needed (see below).

## 5.1 Implications for SE research

While RE research traditionally focuses more on the quality of requirements specifications, the presentation of desiderata also appears important. Presentation issues include not only modeling techniques (e.g. use cases, scenarios, goal models, agent models, IEEE-830 style "the system shall..." statements) but also, as demonstrated here, the language used to convey them. A series of randomized controlled trials and replication studies similar to this study with expert and novice designers as participants are needed to investigate how desiderata framing affects design performance. Qualitative and protocol studies are needed to uncover the cognitive mechanisms underlying these framing effects.

RE research traditionally focuses on prioritizing desiderata and distinguishing mandatory desiderata (i.e. needs) from optional desiderata (i.e. wants). RE may benefit from techniques for indicating the epistemic status of a desideratum, e.g. 80% certainty that the system will need to support encryption; and techniques or tools for representing, analyzing and addressing ill-structured design tasks. We recommend presenting less certain and less important desiderata in a manner that promotes skepticism and is appropriate to the particular context. However, this raises numerous questions for future research including how combining confidence and importance metadata affect creativity and practicality. Moreover, we wonder about the mixed signals of giving a desideratum low confidence or low importance and still labeling it a requirement.

## 5.2 Implications for SE practice

Our results suggest that presenting desiderata as mandatory curtails creativity, independent of the desiderata themselves. If more original solutions are preferred, desiderata should be framed less formally; maybe even to induce skepticism. However, modeling desiderata more formally might result in solutions that are more practical.

While we used a list-of-ideas framing in both the experiments, we do not advocate simply renaming requirements to ideas. The ideas framing is neither supposed to be a realistic approach, nor represent a specific technique in RE. Rather, we encourage professionals to consider which way of representing desiderata—user stories, personas, scenarios, use cases, etc.—is most appropriate for the project at hand. The aim of this study is not to get designers ignore the real requirements, but to encourage them to critically evaluate and recognize real requirements from the dubious ones to produce more creative solutions.

Furthermore, professionals might consider two properties of each desideratum—importance [87] and confidence [88], [89]. While, importance refers to how crucial a desideratum is for success, confidence refers to the certainty of the desideratums relevance. Labeling low-importance, low-confidence desiderata as requirements is likely more problematic than labeling high-importance, high-confidence desiderata as requirements. Here, however, confidence refers to the amount of evidence supporting a desideratum, not how adamantly an individual insists on it.

Perhaps, to promote creativity and innovation, the term *requirement* should be reserved for desiderata that have high importance and ample supporting evidence.

## 5.3 Implications for SE education

Software engineering education continues to advocate oversimplified views of RE and design. The notion that analysts elicit requirements and designers translate those requirements into a system design is simply misleading. SE education should encourage a better understanding of the psychological underpinning of requirements analysis, where an analyst understands requirements as being invented or co-constructed along with stakeholders, and not discovered or elicited in an objective reality as implied by the rationalist philosophy. Moreover, RE vernacular obscures the disagreements and ambiguity in software projects, leading to inaccurate models of desiderata. The results of this study suggest several potential improvements to SE education:

1) More training in creativity enhancing techniques and approaches.
2) More ambiguous projects and ill-structured problems; fewer formal specifications.
3) More exposure to theories and techniques appropriate for ambiguous contexts with conflicting stakeholders, such as actor-network theory [90] and soft systems methodology [91].

## 6 THREATS TO VALIDITY

The two experiments reported here have several limitations. Regarding external validity, participants were not randomly selected from a population; therefore, statistical generalization of results is not possible. Since we targeted experienced but non-expert professionals, the results may not generalize to experts or raw novices. Moreover, because we intentionally compiled a somewhat disorganized list of desiderata, our results may not generalize to highly refined specifications. Since both the experiments reported in this paper concerned a specific type of system (i.e. user-interface of a mobile app), our results may not generalize to other high-compliance or mission-critical systems that are bound by additional constraints such as government policies, legal policies, technical limitations or fixed-contracts, where creativity is achieved by leveraging such mandatory requirements (e.g. aeronautical software). Future replications should involve designing such constrained systems. Additionally, the artificial laboratory setting of the study may have some unknown effects.

Regarding construct validity, originality and practicality are constructs with no objective scales, so we have to measure them using expert judgment. Different judges might produce different results. However, we mitigated this threat by establishing common assessment guidelines, piloting the assessment procedure and asking a third expert judge to resolve disagreements. High inter-rater agreement suggests that, while subjective, the measurement strategy is reliable. That said, originality and practicality are not the only dimensions of design performance, and better designs do not automatically lead to better implementations. Moreover, we only evaluated the creativity of the solution designs (i.e. *product*) but did not assess the creativity potential of the participants. This could be done by using a simplified version of Torrance Test of Creative Thinking [92]. However, research shows that such assessment tests fail to capture all aspects of an individual's creative potential [93].

The controlled nature of the study suggests high internal validity. Similarly, the description of experimental protocol above and our replication package contribute to strong reproducibility. Meanwhile, we used well-understood statistical tests on data the meets their assumptions, contributing to high conclusion validity.

In contrast, our exploratory findings must be interpreted with much more caution. Our manipulation check was implemented as a non-validated, single-item scale, which may not accurately reflect fixation. Because some of the data was non-normal with unequal variances, we had to use multiple, imperfectly suited tests, and therefore did not correct for multiple comparisons. Moreover, combining results across the two studies introduces an unequal groups threat to validity that can only be mitigated with further experiments. In hindsight, we would have ideally preferred one experiment with one control group and two treatment groups. However, we were not confident that the statistical power of such an experiment would be high enough based on the number of available participants, so we broke the study into two separate experiments. Any future replications should consider combining the three groups in a single experiment. Finally, we did not plan these analyses in advance, which invites hypothesizing-after-results-are-known (HARKing). We therefore present these findings as 'exploratory' to emphasize their different epistemic status from our main findings. Despite these limitations, however, we felt that the exploratory findings were sufficiently interesting and relevant to warrant inclusion.

## 7 CONCLUSION

The purpose of this research was to investigate the question—*Does the framing of desiderata affect design creativity?* We conducted two controlled experiments and found that framing desiderata as requirements or prioritized requirements significantly affects design creativity. The major contributions of this paper are two-fold:

1) Presenting desiderata as requirements results in designs that are less original but more practical.
2) Designers are highly susceptible to minor changes in the vernacular used to communicate desiderata.

Contributing to previous research on fixation, we propose the concept of *requirements fixation*—the tendency to rely heavily on desiderata framed as requirements. While previous research in design demonstrated that designers fixate on the features of given examples, our results suggest that designers may also fixate on given desiderata. Minor changes in the vernacular used to communicate the desiderata can manipulate the designers' cognitive functioning, leading to significantly different outcomes. Moreover, the findings emphasize the need for designers to not rely heavily on requirements, and critically evaluate the desiderata presented as requirements to produce more creative solutions. *Requirements fixation* may be mitigated by presenting desiderata more ambiguously, informally and uncertainly.

Future work may investigate requirements fixation more directly, not only by observing professionals in the field but also by developing instruments to measure fixation in the lab. Desiderata can be presented in many formats, and more direct empirical comparisons of these formats are needed.

To conclude, this paper highlights the innate tension between creativity and software requirements. It reveals that the way a specification is presented might be just as important as its contents.

## REFERENCES

[1] T. Chow and D. B. Cao, "A survey study of critical success factors in agile software projects," *Journal of systems and software*, vol. 81, no. 6, pp. 961–971, 2008.

[2] J. Mund, D. M. Fernandez, H. Femmer, and J. Eckhardt, "Does quality of requirements specifications matter? combined results of two empirical studies," in *Empirical Software Engineering and Measurement (ESEM), 2015 ACM/IEEE International Symposium on*. IEEE, 2015, pp. 1–10.

[3] M. Daneva, E. Van Der Veen, C. Amrit, S. Ghaisas, K. Sikkel, R. Kumar, N. Ajmeri, U. Ramteerthkar, and R. Wieringa, "Agile requirements prioritization in large-scale outsourced system projects: An empirical study," *Journal of systems and software*, vol. 86, no. 5, pp. 1333–1353, 2013.

[4] P. Achimugu, A. Selamat, R. Ibrahim, and M. N. Mahrin, "A systematic literature review of software requirements prioritization research," *Information and software technology*, vol. 56, no. 6, pp. 568–585, 2014.

[5] M. Dabbagh, S. P. Lee, and R. M. Parizi, "Functional and non-functional requirements prioritization: empirical evaluation of ipa, ahp-based, and ham-based approaches," *Soft Computing*, vol. 20, no. 11, pp. 4497–4520, 2016.

[6] H. A. Simon, *The sciences of the artificial*. MIT press, 1996.

[7] P. Ralph and R. Mohanani, "Is requirements engineering inherently counterproductive?" in *Proceedings of the Fifth International Workshop on Twin Peaks of Requirements and Architecture*. IEEE Press, 2015, pp. 20–23.

[8] N. Cross, "Expertise in design: an overview," *Design studies*, vol. 25, no. 5, pp. 427–441, 2004.

[9] P. Ralph, "Software engineering process theory: A multi-method comparison of sensemaking–coevolution–implementation theory and function–behavior–structure theory," *Information and Software Technology*, vol. 70, pp. 232–250, 2016.

[10] K. E. Stanovich, *What intelligence tests miss: The psychology of rational thought*. Yale University Press, 2009.

[11] IEEE Computer Society. Software Engineering Standards Committee and IEEE-SA Standards Board, "IEEE recommended practice for software requirements specifications," Institute of Electrical and Electronics Engineers, standard, 1998.

[12] K. Schwaber, *Agile project management with Scrum*. Microsoft press, 2004.

[13] A. Cockburn, "Writing effective use cases, the crystal collection for software professionals," *Addison-Wesley Professional Reading*, 2000.

[14] M. A. Runco and G. J. Jaeger, "The standard definition of creativity," *Creativity Research Journal*, vol. 24, no. 1, pp. 92–96, 2012.

[15] R. Mohanani, P. Ralph, and B. Shreeve, "Requirements fixation," in *Proceedings of the 36th International Conference on Software Engineering*. ACM, 2014, pp. 895–906.

[16] A. Tversky and D. Kahneman, "Judgment under uncertainty: Heuristics and biases," in *Utility, probability, and human decision making*. Springer, 1975, pp. 141–162.

[17] C. Mair and M. Shepperd, "Human judgement and software metrics," in *Proceeding of the 2nd international workshop on Emerging trends in software metrics - WETSoM '11*. New York, New York, USA: ACM Press, 2011, p. 81.

[18] G. Calikli and A. Bener, "Empirical analyses of the factors affecting confirmation bias and the effects of confirmation bias on software developer/tester performance," in *Proceedings of the 6th International Conference on Predictive Models in Software Engineering*. ACM, 2010, p. 10.

[19] N. Chotisarn and N. Prompoon, "Forecasting software damage rate from cognitive bias in software requirements gathering and specification process," in *2013 IEEE Third International Conference on Information Science and Technology (ICIST)*. IEEE, mar 2013, pp. 951–956.

[20] M. Jorgensen and S. Grimstad, "Software Development Estimation Biases: The Role of Interdependence," *IEEE Transactions on Software Engineering*, vol. 38, no. 3, pp. 677–693, may 2012.

[21] W. Stacy and J. MacMillan, "Cognitive bias in software engineering," *Communications of the ACM*, vol. 38, no. 6, pp. 57–63, 1995.

[22] G. J. Browne and V. Ramesh, "Improving information requirements determination: a cognitive perspective," *Information & Management*, vol. 39, no. 8, pp. 625–645, 2002.

[23] K. Mohan and R. Jain, "Using traceability to mitigate cognitive biases in software development," *Communications of the ACM*, vol. 51, no. 9, pp. 110–114, 2008.

[24] J. Parsons and C. Saunders, "Cognitive heuristics in software engineering applying and extending anchoring and adjustment to artifact reuse," *IEEE Transactions on Software Engineering*, vol. 30, no. 12, pp. 873–888, 2004.

[25] R. Hastie and R. M. Dawes, *Rational choice in an uncertain world: The psychology of judgment and decision making*. Sage, 2010.

[26] A. Tversky and D. Kahneman, "The framing of decisions and the psychology of choice," in *Environmental Impact Assessment, Technology Assessment, and Risk Analysis*. Springer, 1985, pp. 107–129.

[27] P. Bohm and H. Lind, "A note on the robustness of a classical framing result," *Journal of Economic Psychology*, vol. 13, no. 2, pp. 355–361, 1992.

[28] P. Ralph, "Toward a theory of debiasing software development," in *EuroSymposium on Systems Analysis and Design*. Springer, 2011, pp. 92–105.

[29] Plous and Scott, *The psychology of judgment and decision making*. Mcgraw-Hill Book Company, 1993.

[30] D. G. Jansson and S. M. Smith, "Design fixation," *Design studies*, vol. 12, no. 1, pp. 3–11, 1991.

[31] R. J. Youmans and T. Arciszewski, "Design fixation: Classifications and modern methods of prevention," *AI EDAM*, vol. 28, no. 2, pp. 129–137, 2014.

[32] R. A. Finke, "Imagery, creativity, and emergent structure," *Consciousness and cognition*, vol. 5, no. 3, pp. 381–393, 1996.

[33] D. Zahner, J. V. Nickerson, B. Tversky, J. E. Corter, and J. Ma, "A fix for fixation? rerepresenting and abstracting as creative processes in the design of information systems," *AI EDAM*, vol. 24, no. 2, pp. 231–244, 2010.

[34] A. T. Purcell and J. S. Gero, "Design and other types of fixation," *Design studies*, vol. 17, no. 4, pp. 363–383, 1996.

[35] M. Perttula and P. Sipilä, "The idea exposure paradigm in design idea generation," *Journal of Engineering Design*, vol. 18, no. 1, pp. 93–102, 2007.

[36] Z. Lujun, "Design fixation and solution quality under exposure to example solution," in *Computing, Control and Industrial Engineering*

*(CCIE), 2011 IEEE 2nd International Conference on*, vol. 1. IEEE, 2011, pp. 129–132.

[37] E. G. Chrysikou and R. W. Weisberg, "Following the wrong footsteps: fixation effects of pictorial examples in a design problemsolving task." *Journal of Experimental Psychology: Learning, Memory, and Cognition*, vol. 31, no. 5, p. 1134, 2005.

[38] O. Atilola, M. Tomko, and J. S. Linsey, "The effects of representation on idea generation and design fixation: A study comparing sketches and function trees," *Design Studies*, vol. 42, pp. 110–136, 2016.

[39] J. Kim and H. Ryu, "A design thinking rationality framework: Framing and solving design problems in early concept generation," *Human–Computer Interaction*, vol. 29, no. 5-6, pp. 516–553, 2014.

[40] D. N. Perkins and D. N. Perkins, *The mind's best work*. Harvard University Press, 2009.

[41] T. M. Amabile, "A model of creativity and innovation in organizations," *Research in organizational behavior*, vol. 10, no. 1, pp. 123–167, 1988.

[42] J. Guilford, "Three faces of intellect1," *Teaching Gifted Students: A Book of Readings*, p. 7, 1965.

[43] J. A. Plucker and M. C. Makel, "Assessment of creativity," *The Cambridge handbook of creativity*, pp. 48–73, 2010.

[44] R. E. Mayer, "22 fifty years of creativity research," *Handbook of creativity*, vol. 449, 1999.

[45] H. H. Christiaans, "Creativity as a design criterion," *Communication Research Journal*, vol. 14, no. 1, pp. 41–54, 2002.

[46] H. G. Nelson and E. Stolterman, *The design way: Intentional change in an unpredictable world: Foundations and fundamentals of design competence*. Educational Technology, 2003.

[47] M. D. Mumford, "Where have we been, where are we going? taking stock in creativity research," *Creativity research journal*, vol. 15, no. 2-3, pp. 107–120, 2003.

[48] M. Sosa and M. Danilovic, "A structured approach to re-organize for creativity," in *DS 58-3: Proceedings of ICED 09, the 17th International Conference on Engineering Design, Vol. 3, Design Organization and Management, Palo Alto, CA, USA, 24.-27.08. 2009*, 2009.

[49] N. Maiden, A. Gizikis, and S. Robertson, "Provoking creativity: Imagine what your requirements could be like," *IEEE software*, vol. 21, no. 5, pp. 68–75, 2004.

[50] J. Diedrich, M. Benedek, E. Jauk, and A. C. Neubauer, "Are creative ideas novel and useful?" *Psychology of Aesthetics, Creativity, and the Arts*, vol. 9, no. 1, p. 35, 2015.

[51] B. T. Christensen and L. J. Ball, "Dimensions of creative evaluation: Distinct design and reasoning strategies for aesthetic, functional and originality judgments," *Design Studies*, vol. 45, pp. 116–136, 2016.

[52] J. J. Shah, S. M. Smith, and N. Vargas-Hernandez, "Metrics for measuring ideation effectiveness," *Design studies*, vol. 24, no. 2, pp. 111–134, 2003.

[53] M. A. Runco and R. E. Charles, "Judgments of originality and appropriateness as predictors of creativity," *Personality and individual differences*, vol. 15, no. 5, pp. 537–546, 1993.

[54] K. Adams, "The sources of innovation and creativity." *National Center on Education and the Economy (NJ1)*, 2005.

[55] J. Baer and J. C. Kaufman, "Bridging generality and specificity: The amusement park theoretical (apt) model of creativity," *Roeper Review*, vol. 27, no. 3, pp. 158–163, 2005.

[56] L. Nguyen and G. Shanks, "A framework for understanding creativity in requirements engineering," *Information and software technology*, vol. 51, no. 3, pp. 655–662, 2009.

[57] R. Horowitz, "Creative problem solving in engineering design," *PhD. diss., Tel-Aviv University*, 1999.

[58] B. Crawford, C. L. de la Barra, R. Soto, and E. Monfroy, "Agile software engineering as creative work," in *Proceedings of the 5th International Workshop on Co-operative and Human Aspects of Software Engineering*. IEEE Press, 2012, pp. 20–26.

[59] S. Dexter and A. Kozbelt, "Free and open source software (foss) as a model domain for answering big questions about creativity," *Mind & Society*, vol. 12, no. 1, pp. 113–123, 2013.

[60] P. Ralph, "The two paradigms of software development research," *Science of Computer Programming*, 2018.

[61] N. Maiden, S. Jones, K. Karlsen, R. Neill, K. Zachos, and A. Milne, "Requirements engineering as creative problem solving: A research agenda for idea finding," in *Requirements Engineering Conference (RE), 2010 18th IEEE International*. IEEE, 2010, pp. 57–66.

[62] S. K. Saha, M. Selvi, G. Büyükcan, and M. Mohymen, "A systematic review on creativity techniques for requirements engineering," in *Informatics, Electronics & Vision (ICIEV), 2012 International Conference on*. IEEE, 2012, pp. 34–39.

[63] N. Maiden and A. Gizikis, "Where do requirements come from?" *IEEE software*, vol. 18, no. 5, pp. 10–12, 2001.

[64] R. Guindon, "Knowledge exploited by experts during software system design," *International Journal of Man-Machine Studies*, vol. 33, no. 3, pp. 279–304, 1990.

[65] J. Wirth, J. Künsting, and D. Leutner, "The impact of goal specificity and goal type on learning outcome and cognitive load," *Computers in Human Behavior*, vol. 25, no. 2, pp. 299–305, 2009.

[66] R. Vollmeyer and B. D. Burns, "Goal specificity and learning with a hypermedia program." *Experimental Psychology*, vol. 49, no. 2, p. 98, 2002.

[67] J. Moss, K. Kotovsky, and J. Cagan, "The influence of open goals on the acquisition of problem-relevant information." *Journal of Experimental Psychology: Learning, Memory, and Cognition*, vol. 33, no. 5, p. 876, 2007.

[68] A. B. Butler, L. L. Scherer, and R. Reiter-Palmon, "Effects of solution elicitation aids and need for cognition on the generation of solutions to ill-structured problems," *Creativity Research Journal*, vol. 15, no. 2-3, pp. 235–244, 2003.

[69] A. T. Bahill and A. M. Madni, "Discovering system requirements," in *Tradeoff Decisions in System Design*. Springer, 2017, pp. 373–457.

[70] V. R. Basili, F. Shull, and F. Lanubile, "Building knowledge through families of experiments," *IEEE Transactions on Software Engineering*, vol. 25, no. 4, pp. 456–473, 1999.

[71] M. Palmiero, R. Nori, V. Aloisi, M. Ferrara, and L. Piccardi, "Domain-specificity of creativity: a study on the relationship between visual creativity and visual mental imagery," *Frontiers in psychology*, vol. 6, p. 1870, 2015.

[72] M. Palmiero and N. Srinivasan, "Creativity and spatial ability: a critical evaluation," *Cognition, Experience and Creativity*, pp. 189–214, 2015.

[73] B. Roskos-Ewoldsen, S. R. Black, and S. M. McCown, "Age-related changes in creative thinking," *The Journal of Creative Behavior*, vol. 42, no. 1, pp. 33–59, 2008.

[74] T. M. Amabile, "The social psychology of creativity: A componential conceptualization." *Journal of personality and social psychology*, vol. 45, no. 2, p. 357, 1983.

[75] J. R. Landis and G. G. Koch, "The measurement of observer agreement for categorical data," *biometrics*, pp. 159–174, 1977.

[76] D. W. Nordstokke and B. D. Zumbo, "A new nonparametric levene test for equal variances." *Psicologica: International Journal of Methodology and Experimental Psychology*, vol. 31, no. 2, pp. 401–430, 2010.

[77] S. S. Shapiro and M. B. Wilk, "An analysis of variance test for normality (complete samples)," *Biometrika*, vol. 52, no. 3/4, pp. 591–611, 1965.

[78] G. Macbeth, E. Razumiejczyk, and R. D. Ledesma, "Cliff's delta calculator: A non-parametric effect size program for two groups of observations," *Universitas Psychologica*, vol. 10, no. 2, pp. 545–555, 2011.

[79] B. Kitchenham, L. Madeyski, D. Budgen, J. Keung, P. Brereton, S. Charters, S. Gibbs, and A. Pohthong, "Robust statistical methods for empirical software engineering," *Empirical Software Engineering*, vol. 22, no. 2, pp. 579–630, 2017.

[80] S. Xu and V. Rajlich, "Dialog-based protocol: an empirical research method for cognitive activities in software engineering," in *Empirical Software Engineering, 2005. 2005 International Symposium on*. IEEE, 2005, pp. 10–pp.

[81] P. Ralph, "The illusion of requirements in software development," *Requirements Engineering*, vol. 18, no. 3, pp. 293–296, 2013.

[82] S. Lichtenstein and P. Slovic, *The construction of preference*. Cambridge University Press, 2006.

[83] M. Jirotka and J. A. Goguen, *Requirements engineering: social and technical issues*. Academic Press Professional, Inc., 1994.

[84] T. Amabile, "Creativity and innovation in organizations," *Harvard Business School Background Note 396-239*, no. 5, 1996.

[85] J. S. Mueller, S. Melwani, and J. A. Goncalo, "The bias against creativity: Why people desire but reject creative ideas," *Psychological science*, vol. 23, no. 1, pp. 13–17, 2012.

[86] N. Cross, K. Dorst, and N. Roozenburg, "Research in design thinking," *Proceedings of a Workshop Meeting Held at the Faculty of Industrial Design Engineering*, 1992.

[87] J. Karlsson, "Software requirements prioritizing," in *Requirements Engineering, 1996., Proceedings of the Second International Conference on*. IEEE, 1996, pp. 110–116.

[88] J. Marchant, C. Tjortjis, and M. Turega, "A metric of confidence in requirements gathered from legacy systems: two industrial case studies," in *Software Maintenance and Reengineering, 2006. CSMR 2006. Proceedings of the 10th European Conference on*. IEEE, 2006, pp. 7–pp.

[89] A. J. Nolan, S. Abrahão, P. Clements, and A. Pickard, "Managing requirements uncertainty in engine control systems development," in *Requirements Engineering Conference (RE), 2011 19th IEEE International*. IEEE, 2011, pp. 259–264.

[90] J. Law and J. Hassard, *Actor network theory and after*. Blackwell Publishers, 1999.

[91] P. Checkland and J. Poulter, *Learning for action: a short definitive account of soft systems methodology and its use, for practitioners, teachers and students*. John Wiley and Sons Ltd, 2006.

[92] P. Torrance, *Thinking Creatively with Pictures, Booklets A and B; [with] Thinking Creatively with Words, Booklets A and B*. Personnel Press, 1966.

[93] E. Torrance, "Predictive validity of the torrance tests of creative thinking," *The Journal of creative behavior*, vol. 6, no. 4, pp. 236–262, 1972.

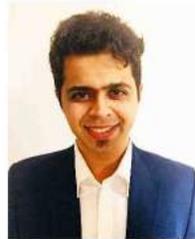

**Rahul Mohanani**, *MSc (Lancaster)*, *B.Eng. (Mumbai)*, is an Asst. Professor in Software Engineering at IIIT, New Delhi, India and a Ph.D. candidate in software engineering at the University of Oulu, Finland. His research focuses on human aspects of software engineering. He has published many peer-reviewed articles in top international conferences and journals, secured multiple scholarships and research grants to support his doctoral studies, and has served on many program committees.

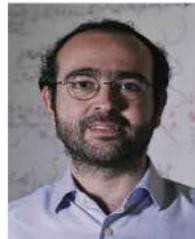

**Burak Turhan**, *PhD (Bogazici)*, is an associate professor in cyber security & systems at Monash University. His research focuses on empirical software engineering, software analytics, quality assurance and testing, human factors, and (agile) development processes. Dr. Turhan has published over 100 articles in international journals and conferences, received several best paper awards, secured funding exceeding 2 million euros, (co-)chaired PROMISE'13, ESEM'17, and PROFES'17, and served on over 30 program committees and editorial boards. For more information please visit: https://turhanb.net.

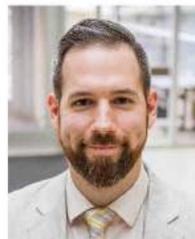

**Paul Ralph**, *PhD (British Columbia)*, *B.Sc. / B.Comm (Memorial)*, is an award-winning scientist, author, consultant and senior lecturer in computer science at the University of Auckland. His research centers on empirical software engineering, human-computer interaction and project management. He has published more than 50 research articles in premier software engineering outlets, won several best paper awards, received funding from Google and The National Sciences and Engineering Research Council of Canada, and served on numerous program committees and editorial boards including ICSE, ESEM and EASE. For more information please visit: http://paulralph.name.